\documentclass[12pt]{article}
\pdfoutput=1
\usepackage{jheppub}
\usepackage{ytableau}
\usepackage{comment}
\usepackage{enumitem}
\usepackage{subfigure}
\usepackage[vcentermath]{youngtab}

\newcommand\be{\begin{equation}}
\newcommand\ee{\end{equation}}

\newcommand\Tr{\mathrm{Tr}}

\newtheorem{thr}{Theorem}

\preprint{KIAS-P22018}

\title{M2-branes and plane partitions}
\abstract{
There is a correspondence between the protected local operators in the 3d SCFTs describing the geometry $\mathbb{C}^2$ 
probed by a stack of $N$ M2-branes and plane partitions of trace $N$. 
We give combinatorial expressions of the indices which count the local operators parametrizing $\mathbb{C}^2/\mathbb{Z}_k$ probed by $N$ M2-branes 
in the canonical and grand canonical ensembles in terms of generating functions for plane partitions. 
We derive the asymptotic behaviors of the grand potential in the high-temperature limit and the scaling dimension in the large $N$ limit. 
}
\author{Tadashi Okazaki}
\emailAdd{tokazaki@kias.re.kr}
\affiliation{
School of Physics, Korea Institute for Advanced Study,\\
85 Hoegi-ro, Cheongnyangri-dong, Dongdaemun-gu, Seoul 02455, Republic of Korea}
\begin{document}
\maketitle

\section{Introduction}


Three-dimensional superconformal field theories (SCFTs) which describe M2-branes have various UV descriptions. 
A stack of $N$ M2-branes moving in $\mathbb{C}^4/\mathbb{Z}_k$ and in $\mathbb{C}^2\times \mathbb{C}^2/\mathbb{Z}_k$ can be described by 
the $U(N)\times U(N)$ ABJM theory with Chern-Simons level $k$ \cite{Aharony:2008ug} 
and by the $U(N)$ ADHM theory with $k$ flavors \cite{Benini:2009qs,Bashkirov:2010kz} respectively. 

They are holographically dual to the eleven-dimensional M-theory 
whose geometry is realized by certain protected local operators on the moduli space of supersymmetric vacua of the SCFTs. 
The ABJM theory has a gauge group $U(N)\times U(N)$ and two types of matter multiplets, a hypermultiplet and a twisted hypermultiplet transforming as the bifundamental representation under the gauge group, which form gauge invariant BPS operators dressing monopole operators. 
The two factors $\mathbb{C}^2/\mathbb{Z}_k$ in the probed eight-dimensional geometry are associated to two kinds of branches of vacua, 
either of which is parametrized by monopoles only dressed by the hypermultiplet or those only dressed by the twisted hypermultiplet. 
On the other hand, the $U(N)$ ADHM theory is 3d $\mathcal{N}=4$ $U(N)$ gauge theory coupled to a single adjoint hypermultiplet and $k$ fundamental hypermultiplets. 
There are protected local operators living on the $\mathcal{N}=4$ Coulomb branch, Higgs branch and mixed branch. 
The geometry $\mathbb{C}^2/\mathbb{Z}_k$ probed by M2-branes is described by the Coulomb branch operators \cite{Mezei:2013gqa}. 

In this paper we examine certain supersymmetric indices for the 3d SCFTs 
which count the local operators realizing the geometry $\mathbb{C}^2/\mathbb{Z}_k$ probed by $N$ M2-branes. 
We obtain closed expressions for the indices in canonical and grand canonical ensembles. 
The result shows a combinatorial nature of the indices as they are expressed in terms of generating functions for plane partitions. 
For $k=1$ we show an exact correspondence between the local operators describing $\mathbb{C}^2$ probed by $N$ M2-branes 
and plane partitions of trace $N$. 
We also obtain closed expressions for the indices with $k=1$ and $2$ in the large $N$ limit. 
It follows that the number of the operators parametrizing $\mathbb{C}^2$ and $\mathbb{C}^2/\mathbb{Z}_2$ probed by M2-branes 
have the asymptotic growth $\sim$ $\mathrm{exp}(\alpha \Delta^{2/3})$ at large scaling dimension $\Delta$ where $\alpha$ is some constant. 
We find that the grand potential in the high-temperature limit near roots of unity has a leading trilogarithm term. 
From the theory of random plane partitions we derive the large scaling dimension of the local operators describing the $\mathbb{C}^2$ probed by $N$ M2-branes. 
It is proportional to $N^{3/2}$ in the large $N$ limit. 

The organization of the paper is as follows. 
In section \ref{sec_canonical} we study the $N$-dependent index, the canonical index for the 3d SCFTs describing the $N$ M2-branes in $\mathbb{C}^2/\mathbb{Z}_k$.  
We obtain a closed formula for the index in terms of generating functions for column-strict plane partitions. 
From the large $N$ indices we also derive the asymptotic growth of the numbers of the operators. 
In section \ref{sec_grand} we examine the grand canonical index which leads to an exact correspondence between 
the local operators parametrizing $\mathbb{C}^2$ probed by $N$ M2-branes and plane partitions of trace $N$. 
We argue that the relation to the column-strict plane partitions follows from the Frobenius construction. 
We also construct the grand potential in the high-temperature limit and derive the scaling dimension of the local operators describing $\mathbb{C}^2$ in the large $N$ limit. 

\section{Canonical index}
\label{sec_canonical}
\subsection{Definition}

We consider the 3d SCFTs equipped with the chiral algebra $\mathcal{A}$ formed by the local operators of scaling dimension $\Delta$ on the moduli space of supersymmetric vacua. 
For such theories we can define a supersymmetric index 
\begin{align}
\label{index}
\mathcal{I} [\mathcal{T} ] (q)&={\Tr}_{\mathcal{A}} (-1)^F q^{2 \Delta}
\end{align}
as a graded character of $\mathcal{A}$ where $F$ is the Fermion number operator. 
The index (\ref{index}) can be obtained from the flavored superconformal index \cite{Kim:2009wb,Imamura:2011su,Kapustin:2011jm} as a special fagacity limit \cite{Razamat:2014pta}. 

In the following we take $\mathcal{A}$ as the chiral algebra formed by the local operators which describe $\mathbb{C}^2/\mathbb{Z}_k$ probed by M2-branes. 
There are several UV descriptions of the index (\ref{index}). 
For example, for the $U(N)\times U(N)$ ABJM theory with Chern-Simons level $k$ describing $N$ M2-branes in $\mathbb{C}^4/\mathbb{Z}_k$, 
the index (\ref{index}) enumerates the gauge invariant local operators which are responsible for the $\mathbb{C}^2/\mathbb{Z}_k$ consisting of monopole operators only dressed by twisted hypermultiplet scalars or equivalently those only dressed by hypermultiplet scalars \cite{Hayashi:2022ldo}.  
On the other hand, for the $U(N)$ ADHM theory with $k$ flavors describing $N$ M2-branes in $\mathbb{C}^2\times \mathbb{C}^2/\mathbb{Z}_k$, 
the $\mathbb{C}^2/\mathbb{Z}_k$ is parametrized by the $\mathcal{N}=4$ Coulomb branch operators \cite{Mezei:2013gqa}. 
Thus the index (\ref{index}) simultaneously counts the Coulomb branch operators in the ADHM theory. 
It can be evaluated by employing the formula proposed in \cite{Cremonesi:2013lqa}. 
Hence we get 
\footnote{
More generally, this also counts the gauge invariant operators in the mirror necklace quiver theories \cite{deBoer:1996mp,deBoer:1996ck}
and those in the $\mathcal{N}=4$ circular quiver Chern-Simons matter theories \cite{Imamura:2008ji,Imamura:2008dt}
which describe the geometry $\mathbb{C}^2/\mathbb{Z}_k$ probed by $N$ M2-branes. 
The index (\ref{c2zk_index}) follows from $\mathcal{N}=4$ flavored superconformal index \cite{Hayashi:2022ldo}. 
}
\begin{align}
\label{c2zk_index}
\mathcal{I}_{N,k}(q)&=\mathcal{I} [\textrm{$U(N)_k\times U(N)_{-k}$ ABJM}] (q)
\nonumber\\
&=\mathcal{I}[\textrm{$U(N)$ ADHM with $k$ flavors}] (q)
\end{align}
by choosing $\mathcal{A}$ as the chiral algebra formed by the local operators which describe $\mathbb{C}^2/\mathbb{Z}_k$ probed by $N$ coincident M2-branes.
We refer to (\ref{c2zk_index}) as the canonical index as it depends on the number $N$ of M2-branes. 
It is the generating function, a.k.a. Hilbert series for chiral operators on the branch of supersymmetric vacua in the M2-brane SCFT 
which are responsible for $\mathbb{C}^2/\mathbb{Z}_k$ in M-theory. 

In this section we explicitly explain how the canonical index (\ref{c2zk_index}) can be evaluated in terms of generating functions for plane partitions 
and how it counts the local operators in the M2-brane SCFTs. 

\subsection{Combinatorial formula}
We begin by introducing basic notions of plane partitions (see e.g. \cite{MR325407,MR1634067}). 
A plane partition of $n$ is an array of non-negative integers
\begin{align}
\label{pp}
\begin{array}{cccc}
n_{11}&n_{12}&n_{13}&\cdots \\
n_{21}&n_{22}&n_{23}&\cdots \\
\vdots&\vdots&\vdots \\
\end{array}
\end{align}
with $\sum_{i,j}n_{ij}=n$. 
The rows and columns in (\ref{pp}) are arranged in non-increasing order. 
The non-zero entries $n_{ij}>0$ are called the parts of the plane partition, 
the sum $n=\sum_{i,j}n_{ij}$ of all entries is called the norm of the plane partition, 
the sum $\sum_{i}n_{ii}$ of the diagonal entries is called the trace of the plane partition. 
Let $\lambda_i$ be parts in the $i$-th row of the plane partition 
in such a way that 
\begin{align}
\lambda_1\ge \lambda_2\ge\cdots \ge \lambda_{r}>\lambda_{r+1}=0. 
\end{align}
The partition $\lambda$ is called the shape of the plane partition. 
If the entries in the plane partition strictly decrease in each column, it is called column-strict. 

Let $\alpha_{\lambda}(n)$ be the number of column-strict plane partitions of $n$ of shape $\lambda$ with $N=\sum_i \lambda_i$. 
One can define a generating function for $\alpha_{\lambda}(n)$ by
\begin{align}
\label{block}
\chi_{\lambda}(q)&=\sum_{n=0}^{\infty} \alpha_{\lambda}(n) q^{n-N}. 
\end{align}
It is given by \cite{MR325407,MR2213154,MR309748}
\begin{align}
\label{block_IR}
\chi_{\lambda}(q)&=
\prod_{b\in \lambda}
\frac{q^{n_{\lambda}-N}}{1-q^{h_{\lambda}(b)}}
\end{align}
where 
\begin{align}
n_{\lambda}&=\sum_{i=1}^{c}
\left(
\begin{matrix}
\lambda_i'+1\\
2\\
\end{matrix}
\right)=\sum_{i=1}^{r} i \lambda_i 
\end{align}
and 
\begin{align}
h_{\lambda}(b)&=
\lambda_i+\lambda_j'-i-j+1
\end{align}
is the hook length of a box $b$ at the $i$-th row and $j$-th column in the corresponding Young diagram $\lambda$. 
$\lambda_i'$ is the number of boxes in the $i$-th column of the Young diagram $\lambda$ 
and $c$ is the number of columns. 

The expression (\ref{block_IR}) is a classical limit of the Verma character \cite{Gaiotto:2019mmf}
for the quantized algebra known as the spherical part of the rational Cherednik algebra \cite{Kodera:2016faj}. 
So the generating function $\chi_{\lambda}(q)$ is identified with the character of the $N$-th symmetric product algebra. 

For $k=1$ one finds that the index (\ref{c2zk_index}) can be simply expressed in terms of the generating function (\ref{index}) as
\begin{align}
\label{SymNC2}
\mathcal{I}_{N,k=1}(q)&=
\sum_{\lambda} 
\chi_{\lambda}(q)^2
\end{align}
where the sum is taken over the Young diagram $\lambda$ with $N=\sum_i \lambda_i$. 
The Young diagram corresponds to the isolated massive vacua of the UV gauge theory 
in which certain operators get vevs. 

For example, when $N=4$ and $k=1$ we have five terms labeled by the Young diagrams
\begin{align}
\chi_{\tiny \yng(4)}(q)&=\frac{1}{(1-q)(1-q^2)(1-q^3)(1-q^4)},\nonumber\\ 
\chi_{\tiny \yng(3,1)}(q)&=\frac{q}{(1-q)^2(1-q^2)(1-q^4)}, \nonumber\\
\chi_{\tiny \yng(2,2)}(q)&=\frac{q^2}{(1-q)(1-q^2)^2(1-q^3)}, \nonumber\\
\chi_{\tiny \yng(2,1,1)}(q)&=\frac{q^3}{(1-q)^2(1-q^2)(1-q^4)}, \nonumber\\
\chi_{\tiny \yng(1,1,1,1)}(q)&=\frac{q^6}{(1-q)(1-q^2)(1-q^3)(1-q^4)}, 
\end{align}
Summing over the Young diagrams we get 
\begin{align}
\label{Sym4C2}
\mathcal{I}_{4,1}(q)&=
\chi_{\tiny \yng(4)}(q)^2
+\chi_{\tiny \yng(3,1)}(q)^2
+\chi_{\tiny \yng(2,2)}(q)^2
+\chi_{\tiny \yng(2,1,1)}(q)^2
+\chi_{\tiny \yng(1,1,1,1)}(q)^2
\nonumber\\
&=
\frac{
1+q^2+2q^3+4q^4+2q^5+4q^6+2q^7+4q^8+2q^9+q^{10}+q^{12}
}
{(1+q)^4(1+q^2)^2(1+q+q^2)^2(1-q)^8}
\nonumber\\
&=1+2q+6q^2+14q^3+33q^4+64q^5+127q^6+228q^7+404q^8+672q^9+\cdots
\end{align}
It can be checked that (\ref{Sym4C2}) correctly counts the gauge invariant protected operators describing $\mathrm{Sym}^4(\mathbb{C}^2)$ 
from the $U(4)_1\times U(4)_{-1}$ ABJM theory or the $U(4)$ ADHM theory with one flavor. 

In the $U(4)_1\times U(4)_{-1}$ ABJM theory 
the monopole operators carry an electric charge because of the Chern-Simons term. 
The Gauss law implies that the gauge invariant monopole operator which contributes to the index has no scaling dimension. 
The term $2q$ in (\ref{Sym4C2}) counts two kinds of dressed monopole $v^{1,0,0,0;1,0,0,0}T$ and $v^{-1,0,0,0;-1,0,0,0}\tilde{T}$ 
where $v^{\{m_i\};\{m_i\}}$ is the bare monopole with magnetic fluxes $\{m_i\}_{i=1}^{4}$, 
and $(T,\widetilde{T})$ is the bifundamental twisted hyper. 
The term $6q^2$ in (\ref{Sym4C2}) corresponds to the following six gauge invariant operators: 
\begin{align}
&\Tr (T\widetilde{T}),& 
&v^{1,-1,0,0;1,-1,0,0}T\widetilde{T},
\nonumber\\
&v^{2,0,0,0;2,0,0,0}T^2,& 
&v^{1,1,0,0;1,1,0,0}T^2,
\nonumber\\
&v^{-2,0,0,0;-2,0,0,0}\widetilde{T}^2,& 
&v^{-1,-1,0,0;-1,-1,0,0}\widetilde{T}^2. 
\end{align}

Unlike the ABJM model, in the $U(4)$ ADHM theory with one flavor the bare monopole of magnetic flux $\{m_i\}_{i=1}^{4}$ has dimension $\sum_i |m_i|/2$ \cite{Gaiotto:2008ak}. 
It is a gauge invariant operator by itself. 
The term $2q$ corresponds to the two fundamental bare monopoles $v^{1,0,0,0}$ and $v^{-1,0,0,0}$. 
The term $6q^2$ enumerates six Coulomb branch operators
\begin{align}
&\Tr \varphi,& 
&v^{1,-1,0,0},\nonumber\\
&v^{2,0,0,0}& 
&v^{1,1,0,0},\nonumber\\
&v^{-2,0,0,0}& 
&v^{-1,-1,0,0}. 
\end{align}
where $\varphi$ is the vector multiplet scalar field. 
The order $8$ of the pole at $q=1$ in the index (\ref{Sym4C2}) is equal to the complex dimension of $\mathrm{Sym}^4(\mathbb{C}^2)$. 

The combinatorial formula (\ref{SymNC2}) can be generalized to the case with $k>1$ where the $N$ M2-branes propagate in $\mathbb{C}^2/\mathbb{Z}_k$. 
Let 
\begin{align}
\label{sequence}
0\le N_1\le N_2\le \cdots\le N_{k-1}\le N
\end{align}
be a non-decreasing sequence of integers. 
Let $\lambda^{(1)}$, $\lambda^{(2)}$, $\cdots$, $\lambda^{(k)}$ be $k$ partitions whose weights are given by 
\begin{align}
\label{set_Young}
\sum_{i}\lambda^{(1)}_{i}&=N_1, \nonumber\\
\sum_{i}\lambda^{(2)}_i&=N_2-N_1, \nonumber\\
\sum_{i}\lambda^{(3)}_i&=N_3-N_2, \nonumber\\
&\vdots \nonumber\\
\sum_{i}\lambda^{(k)}_i&=N-N_{k-1}. 
\end{align}
We find that the index (\ref{c2zk_index}) for general $N$ and $k$ can be expressed in terms of the generating function (\ref{block}) for column-strict plane partitions as 
\begin{align}
\label{SymNC2Zk}
\mathcal{I}_{N,k}(q)&=
\sum_{0\le N_1\le\cdots\le N_{k-1}\le N}
\sum_{\lambda^{(1)}, \cdots, \lambda^{(k)}} \chi_{\lambda^{(1)}}(q^{k})^2
\cdots 
\chi_{\lambda^{(k)}}(q^{k})^2
q^{2(k-1)N-2\sum_{i=1}^{k-1}N_i}
\end{align}
where the sum is first taken over the $k$ sets of Young diagrams characterized by the partitions (\ref{set_Young}) for the fixed sequence (\ref{sequence}) 
and then over the sequences (\ref{sequence}). 

The formula (\ref{SymNC2Zk}) can be also expressed in terms of (\ref{SymNC2}). 
Thus we can write 
\begin{align}
\label{SymNC2Zk_sub}
\mathcal{I}_{N,k}(q)&=
\sum_{0\le N_1\le\cdots\le N_{k-1}\le N}
\mathcal{I}_{N_1,1}(q^k)
\cdots 
\mathcal{I}_{N-N_{k-1},1}(q^k)
q^{2(k-1)N-2\sum_{i=1}^{k-1}N_i}
\end{align}
where we have defined $\mathcal{I}_{0,1}(q)=1$. 

For example, when $N=2$ and $k=4$ we have
\begin{align}
\label{Sym2Z4}
\mathcal{I}_{2,4}(q)&=
\underbrace{\mathcal{I}_{2,1}(q^4)q^{12}}_{(N_1,N_2,N_3)=(0,0,0)}
+\underbrace{\mathcal{I}_{1,1}(q^4)^2q^{10}}_{(N_1,N_2,N_3)=(0,0,1)}
+\underbrace{\mathcal{I}_{1,1}(q^4)^2q^8}_{(N_1,N_2,N_3)=(0,1,1)}
+\underbrace{\mathcal{I}_{1,1}(q^4)^2q^6}_{(N_1,N_2,N_3)=(1,1,1)}
\nonumber\\
&+\underbrace{\mathcal{I}_{2,1}(q^4)q^8}_{(N_1,N_2,N_3)=(0,0,2)}
+\underbrace{\mathcal{I}_{1,1}(q^4)^2q^6}_{(N_1,N_2,N_3)=(0,1,2)}
+\underbrace{\mathcal{I}_{1,1}(q^4)^2q^4}_{(N_1,N_2,N_3)=(1,1,2)}
\nonumber\\
&+\underbrace{\mathcal{I}_{2,1}(q^4)q^4}_{(N_1,N_2,N_3)=(0,2,2)}
+\underbrace{\mathcal{I}_{1,1}(q^4)^2q^2}_{(N_1,N_2,N_3)=(1,2,2)}
\nonumber\\
&+\underbrace{\mathcal{I}_{2,1}(q^4)}_{(N_1,N_2,N_3)=(2,2,2)}
\nonumber\\
&=\frac{1-q^2+2q^4+2q^8-q^{10}+q^{12}}{(1+q^2)^2(1+q^4)(1-q^2)^4}
\nonumber\\
&=1+q^2+4q^4+6q^6+14q^8+19q^{10}+33q^{12}+44q^{14}+\cdots
\end{align}
Again we can check that (\ref{Sym2Z4}) reproduces the result obtained from 
the $U(2)\times U(2)$ ABJM theory with level $k=4$ or the $U(2)$ ADHM theory with four flavors. 

The term $q^2$ corresponds to the gauge invariant operator $\Tr (T\widetilde{T})$ of dimension $1$ in the $U(2)_{4}\times U(2)_{-4}$ ABJM theory. 
Due to the Chern-Simons coupling with $k=4$, the bare monopole carries four units of electric charges 
so that the fundamental monopole with $(m_1,m_2)$ $=$ $(1,0)$, $(-1,0)$ can form gauge invariant operators 
when dressed by fourth power of $T$ and $\widetilde{T}$. 
Also the $U(2)$ gauge group allows for single and double trace operators as gauge invariant operators. 
So there exist four gauge invariant operators with four units of canonical R-charge or scaling dimension
\begin{align}
&v^{1,0;1,0}T^4,& 
&v^{-1,0;-1,0}\widetilde{T}^4,
\nonumber\\
&\Tr (T\widetilde{T}T\widetilde{T}),& 
&\Tr(T\tilde{T})\Tr(T\widetilde{T}). 
\end{align}
These correspond to the term $4q^4$ in (\ref{Sym2Z4}). 

On the other hand, in the $U(2)$ ADHM theory with four flavors, the term $q^2$ comes from $\Tr \varphi$. 
The bare monopole in the $U(2)$ ADHM theory with four flavors has dimension $2\sum_{i=1}^{2} |m_i|$ \cite{Gaiotto:2008ak} so that the fundamental monopole $v^{\pm,0}$ has dimension $2$. 
Besides, there are single and double trace operators consisting of $\varphi$. 
Thus the term $4q^4$ counts the following four Coulomb branch operators
\begin{align}
&v^{1,0},& 
&v^{-1,0},
\nonumber\\
&\Tr(\varphi^2),&
&\Tr(\varphi)\Tr(\varphi). 
\end{align}
The order $4$ at pole $q=1$ in (\ref{Sym2Z4}) is the complex dimension of $\mathrm{Sym}^2(\mathbb{C}^2/\mathbb{Z}_4)$. 

We can also write the index (\ref{c2zk_index}) for $k=1$ as
\begin{align}
\label{PN}
\mathcal{I}_{N,1}(q)&=\frac{P_N(q)}{(q;q)_{N}^2}. 
\end{align}
Here $P_N(q)$ is a palindromic polynomial in $q$ with non-negative integer coefficient of degree $N(N-1)$. 
We observe that it satisfies a relation
\begin{align}
\label{PN_R_transf}
P_N(q^{-1})&=q^{-N(N-1)}P_N(q). 
\end{align}
We also observe that 
the terms $q^k$ with $k=0,1,2,\cdots, N$ in $P_N(q)$ are the same as those in the following function: 
\begin{align}
&\prod_{n=1}^{\infty} \frac{1}{(1-q^n)^{n-1}}
\nonumber\\
&=1+q^2+2q^3+4q^4+6q^5+12q^6+18q^7+33q^8+52q^9+\cdots
\end{align}
For example, we have
\begin{align}
P_1(q)&=1,\nonumber\\
P_2(q)&=1+q^2,\nonumber\\
P_3(q)&=1+q^2+2q^3+q^4+q^6,\nonumber\\
P_4(q)&=1+q^2+2q^3+4q^4+2q^5+4q^6+2q^7+4q^8+2q^9+q^{10}+q^{12},\nonumber\\
P_5(q)&=1+q^2+2q^3+4q^4+6q^5+7q^6+8q^7+12q^8+12q^9+14q^{10}
\nonumber\\
&+\mathrm{palindromic}+q^{20}
\end{align}
Another observation is that 
when $q\rightarrow 1$ the polynomial $P_N(q)$ turns into an ordinary factorial
\begin{align}
\label{PN_1}
P_N(1)&=N!. 
\end{align}

From (\ref{PN_R_transf}), (\ref{PN}) and (\ref{SymNC2Zk_sub}) it follows that 
the index (\ref{c2zk_index}) satisfies a relation
\begin{align}
\label{R_transf}
\mathcal{I}_{N,k}(q^{-1})&=q^{2N}\mathcal{I}_{N,k}(q). 
\end{align}
Thus the normalized index $q^N \mathcal{I}_{N,k}(q)$ is invariant under the transformation $q\rightarrow q^{-1}$. 

\subsection{Large $N$ limit}
Now consider the large $N$ limit of the index. 
For $k=1$ the index (\ref{c2zk_index}) has the following expansions: 
\begin{align}
\label{HS_list}
\begin{array}{c|l} \hline\hline 
\textrm{$\#$ (M2-branes)}&\qquad \qquad \qquad \qquad \qquad \textrm{Expansion} \\ \hline
1&1+2 q+3q^2+4q^3+5q^4+6q^5+7q^6+8q^7+9q^8+\cdots \\ 
2&1+2 q+6 q^2+10 q^3+19 q^4+28 q^5+44 q^6+60 q^7+85 q^8+\cdots \\ 
3&1+2 q+6 q^2+14 q^3+28 q^4+52 q^5+93 q^6+152 q^7+242 q^8+\cdots \\ 
4&1+2 q+6 q^2+14 q^3+33 q^4+64 q^5+127 q^6+228 q^7+404 q^8+\cdots \\ 
5&1+2 q+6 q^2+14 q^3+33 q^4+70 q^5+142 q^6+272 q^7+507 q^8+\cdots \\ 
6&1+2 q+6 q^2+14 q^3+33 q^4+70 q^5+149 q^6+290 q^7+561 q^8+\cdots \\ 
7&1+2 q+6 q^2+14 q^3+33 q^4+70 q^5+149 q^6+298 q^7+582 q^8+\cdots \\ 
8&1+2 q+6 q^2+14 q^3+33 q^4+70 q^5+149 q^6+298 q^7+591 q^8+\cdots \\ 
\hline 
\end{array}
\end{align}
As seen from (\ref{HS_list}) the finite-$N$ correction of the index appears from $q^{N+1}$. 
This property is the same as the full supersymmetric index due to the existence of multi-trace operators. 
We have numerically confirmed that the index (\ref{c2zk_index}) for $k=1$ in the large $N$ limit is given by 
\begin{align}
\label{largeNk1}
\mathcal{I}_{\infty,1}(q)&=\prod_{n=1}^{\infty}\frac{1}{(1-q^n)^{n+1}}
\nonumber\\
&=1+2 q+6 q^2+14 q^3+33 q^4+70 q^5+149 q^6+298 q^7+591 q^8
\nonumber\\
&+1122q^9+2101q^{10}+3822q^{11}+6848q^{12}+\cdots
\end{align}
Noticing that (see \cite{MR2445243}) 
\begin{align}
E(q)&:=\prod_{n=1}^{\infty} \frac{1}{1-q^n}=\sum_{n=0}^{\infty} p(n) q^n
\end{align}
and \cite{macmahon1912ix} 
\begin{align}
\label{MacMahon}
M(q)&:=\prod_{n=1}^{\infty} \frac{1}{(1-q^n)^n}=\sum_{n=0}^{\infty} Q(n) q^n
\end{align}
where $p(n)$ is the number of partitions of $n$ and $Q(n)$ is the number of plane partitions of $n$, 
we can write 
\begin{align}
\label{largeNk1_sub}
\mathcal{I}_{\infty,1}(q)&=
\sum_{n=0}^{\infty}\sum_{N=0}^{n} p(N)Q(n-N)q^n. 
\end{align}
This implies that the large $N$ index (\ref{largeNk1}) can be identified with a generating function for plane partitions of $N+n$ with trace $N$ 
which is associated to the partition of a diagonal parts.  
We will see in section \ref{sec_grand} that there is an exact correspondence between the local operators describing the motion of $N$ M2-branes in $\mathbb{C}^2$ 
and the plane partitions of trace $N$ in the analysis of the grand canonical ensemble. 

One aspect which we can obtain from the expression (\ref{largeNk1}) is the asymptotic growth of the number of operators. 
When one writes the 3d index (\ref{index}) for a free scalar theory as an integral over a density $\rho(\Delta)$ of scaling dimension $\Delta$, 
then its asymptotic behavior takes the form \cite{Cardy:1991kr}
\footnote{
For a free scalar theory in $d$ dimensions, 
the growth of the number of operators of scaling dimension $\Delta$ is given by \cite{Cardy:1991kr}
\begin{align}
\label{density_d}
\rho(\Delta)&\sim 
\exp\left(
\alpha \Delta^{1-1/d}
\right).
\end{align}
Also see e.g. \cite{Kutasov:2000td,Henning:2017fpj,Melia:2020pzd} for further studies of the asymptotic growth.  
}
\begin{align}
\label{density_3d}
\rho(\Delta)&\sim \exp\left( \alpha \Delta^{2/3} \right)
\end{align}
where $\alpha$ is some constant. 
Although much less is known for interacting 3d theories, 
we can obtain from (\ref{largeNk1}) the asymptotic growth of the number $a_n$ of the local operators of dimension $\Delta=n/2$ 
in the 3d SCFTs for M2-branes parametrizing $\mathbb{C}^2$ in the large $N$ limit. 
The following theorem by Meinardus \cite{MR62781, MR1634067} holds the key to the asymptotic growth of the number of operators:
\begin{thr}
For an infinite product with the form 
\begin{align}
f(q)&=\prod_{n=1}^{\infty} \frac{1}{(1-q^n)^{b_n}}=1+\sum_{n=1}^{\infty} a_n q^n
\end{align}
where $q=e^{-\beta}$ and $\mathrm{Re}\beta>0$, 
consider an auxiliary Dirichlet series 
\begin{align}
D(s)&=\sum_{n=1}^{\infty}\frac{b_n}{n^s},\qquad s=\sigma+i\tau. 
\end{align}
Suppose that we have the following conditions:

\begin{enumerate}[label=(\roman*)]

\item Condition

 $D(s)$ converges for $\sigma>\alpha$, a positive number and has an analytic continuation in the region $\sigma\ge -C_0$ with $0<C_0<1$, 

\item Condition
 
$D(s)$ is analytic except for a pole of order $1$ at $s=s_0$ with residue $R_{0}$ 

\item Condition

$D(s)$ $\rightarrow$ $\mathcal{O}(|\tau|^{C_1})$ as $|\tau|\rightarrow \infty$ 
for a fixed positive number $C_1$. 

\end{enumerate}
Then we have 
\begin{align}
a_n&\sim 
Cn^{\kappa} \exp 
\left[
n^{\frac{s_0}{s_0+1}}
\left(
1+\frac{1}{s_0} 
\right)
\left(
R_{0}\Gamma(s_0+1) \zeta(s_0+1)
\right)^{\frac{1}{s_0+1}}
\right]& 
&\textrm{as $n\rightarrow \infty$}
\end{align}
where 
\begin{align}
C&=e^{D'(0)}\left[2\pi(1+s_0) \right]^{-\frac12}
\left[
R_{0}\Gamma(s_0+1)\zeta(s_0+1)
\right]^{(1-2D(0))/(2+2s_0)}
\end{align}
and 
\begin{align}
\kappa&=\frac{D(0)-1-\frac{s_0}{2}}{1+s_0}. 
\end{align}
\end{thr}
Since the index (\ref{largeNk1}) leads to the Dirichlet series 
$\sum_{n=1}^{\infty} (n+1)/n^s$ $=$ $\zeta(s-1)+\zeta(s)$ with two poles at $s=1$ and $s=2$, 
the second condition is not satisfied. 
When there exist additional poles at $s=s_i$ $(<$ $s_0)$ in the Dirichlet series, 
the sub-leading terms generally appear and we therefore need a generalization of the Meinardus Theorem. 
The generalized theorem that is applicable for multiple poles is presented in \cite{MR2958955}. 
The main idea based on the auxiliary Dirichlet series remains the same. 
Applying the generalized Meinardus Theorem to the index (\ref{largeNk1}), 
we obtain the asymptotic behavior of the number $a_n$ of the operators
\begin{align}
\label{c2_asy}
a_n
\sim 
C n^{\kappa}
\exp\left(
\alpha n^{2/3}+\beta n^{1/3}+\gamma
\right), \qquad n=2\Delta
\end{align}
where 
\begin{align}
\label{MacMah_alpha}
\alpha&=\frac{3\zeta(3)^{1/3}}{2^{2/3}}=2.00944...,\\
\beta&=\frac{\pi^2}{3\cdot 2^{4/3} \zeta(3)^{1/3}}=1.22790...,\\
\gamma&=\frac{1}{12}-\frac{\pi^4}{432\zeta(3)}=-0.10424..., \\
C&=\frac{\zeta(3)^{13/36}}{2^{23/36}\cdot 3^{1/2}\pi A}=0.098354..., \\
\kappa&=-31/36=-0.86111...
\end{align}
and $A$ is the Glaisher-Kinkelin constant. 
We see that the asymptotic growth (\ref{c2_asy}) takes the same form as (\ref{density_3d}). 
The leading coefficient $\alpha$ given by (\ref{MacMah_alpha}) coincides with that appearing in the asymptotic growth of the MacMahon function 
\cite{MR1575956}. 
The exact numbers $\mathcal{N}(n)$ of the local operators and the values $a_n$ computed from (\ref{c2_asy}) are listed as follows: 
\begin{align}
\label{c2_asy_table}
\begin{array}{c|c|c} 
n&\mathcal{N}(n)&a_n \\ \hline 
10&2139&1931.87 \\
100&3.42106\times 10^{18}&3.17747\times 10^{18} \\
1000&9.63125\times 10^{88}&9.24720\times 10^{88} \\
5000&1.17082\times 10^{260}&1.14167\times 10^{260} \\
10000&1.17013\times 10^{412}&1.14657\times 10^{412} \\
\end{array}
\end{align}

We remark that the large $N$ full superconformal index for M2-branes in flat space is shown to agree with the Kaluza-Klein index $I_{\textrm{KK}}$ \cite{Kim:2009wb}. 
The finite-$N$ full superconformal index $I_N$ for M2-branes which contains contributions from a stack of $N$ M2-branes will take the form \cite{Arai:2020uwd,Gaiotto:2021xce}
\begin{align}
I_{N}(q)&=I_{\textrm{KK}}(q)
\left(
1+\sum_{\textrm{conf.}}I_{\textrm{M5}}(q)
\right)
\end{align}
where $I_{\textrm{M5}}(q)$ is the finite-$N$ correction to the index due to the wrapped M5-branes. 
From (\ref{SymNC2}) and (\ref{largeNk1}) we find the following finite-$N$ correction: 
\begin{align}
\label{finiteN}
\frac{\mathcal{I}_{N,1}(q)}{\mathcal{I}_{\infty,1}(q)}
&=
\prod_{n=1}^{\infty} (1-q^n)^{n+1}
\sum_{\lambda} \chi_{\lambda}(q)^2
\nonumber\\
&=
\frac{(q;q)_{\infty}^2}
{(q;q)_{N}^2}
\prod_{n=1}^{\infty} (1-q^n)^{n-1}P_N(q). 
\end{align}
Although we do not pursue it here, 
it would be interesting to explore a generalization of this relation to 
the full superconformal index and to understand it from the point of view of the M5-branes. 

For $k=2$ we obtain the index 
\begin{align}
\label{HS_list2}
\begin{array}{c|l} \hline\hline 
\textrm{$\#$ (M2-branes)}&\qquad \qquad \qquad \qquad \qquad \textrm{Expansion} \\ \hline
1&1+3q^2+5q^4+7q^6+9q^8+11q^{10}+13q^{12}+15q^{14}+17q^{16}+\cdots \\ 
2&1+3q^2+11 q^4+22 q^6+45 q^8+73 q^{10}+119 q^{12}+172 q^{14}+249q^{16}+\cdots \\ 
3&1+3q^2+ 11q^4+32 q^6+75 q^8+160 q^{10}+313 q^{12}+562 q^{14}+956q^{16}+\cdots \\ 
4&1+3q^2+ 11q^4+32 q^6+90 q^8+210 q^{10}+473 q^{12}+967 q^{14}+1889q^{16}+\cdots \\ 
5&1+3q^2+ 11q^4+32 q^6+90 q^8+231 q^{10}+548 q^{12}+1222 q^{14}+2584q^{16}+\cdots \\ 
6&1+3q^2+ 11q^4+32 q^6+90 q^8+231 q^{10}+576 q^{12}+1327 q^{14}+2956q^{16}+\cdots \\ 
7&1+3q^2+ 11q^4+32 q^6+90 q^8+231 q^{10}+576 q^{12}+1363 q^{14}+3096q^{16}+\cdots \\ 
8&1+3q^2+ 11q^4+32 q^6+90 q^8+231 q^{10}+576 q^{12}+1363 q^{14}+3141q^{16}+\cdots \\ 
\hline 
\end{array}
\end{align}
Again we see that the finite-$N$ correction shows up from $q^{N+1}$. 
In this case we numerically find that the large $N$ index is simply given by 
\begin{align}
\label{largeN_k2}
\mathcal{I}_{\infty,2}(q)&=
\prod_{n=1}^{\infty} \frac{1}{(1-q^{2n})^{2n+1}}. 
\end{align}
Applying the generalized Meinardus Theorem in \cite{MR2958955} to the index (\ref{largeN_k2}), 
we find that the number $a_n$ of the operators has the same asymptotic growth as (\ref{c2_asy}) with 
\begin{align}
\label{z2_alpha}
\alpha&=\frac{3\zeta(3)^{1/3}}{2}=1.59490...,\\
\beta&=\frac{\pi^2}{3\cdot 2^2 \zeta(3)^{1/3}}=0.773529...,\\
\gamma&=\frac{1}{6}-\frac{\pi^4}{364\zeta(3)}=-0.0559579...,\\
C&=\frac{2^{7/9} \zeta(3)^{1/2}}{3^{1/2}\pi A^2}=0.210049...,\\
\label{z2_kappa}
\kappa&=-\frac{8}{9}=-0.888889...
\end{align}
We show the exact numbers $\mathcal{N}(n)$ of the local operators and the analytic values $a_n$ 
evaluated from (\ref{c2_asy}) as well as (\ref{z2_alpha})-(\ref{z2_kappa})
\begin{align}
\label{c2z2_asy_table}
\begin{array}{c|c|c} 
n&\mathcal{N}(n)&a_n \\ \hline 
10&231&222.798 \\
100&1.07823\times 10^{14}&1.00541\times 10^{14} \\
1000&1.91038\times 10^{69}&1.80422\times 10^{69} \\
5000&2.03769\times 10^{204}&1.94181\times 10^{204} \\
10000&3.17566\times 10^{324}&3.03553\times 10^{324} \\
\end{array}
\end{align}
We leave it for future work to present the analytic and numerical treatments of the asymptotic growth for $k>2$. 

\section{Grand canonical index}
\label{sec_grand}

\subsection{Operators and plane partitions}
An alternative approach is to consider the grand canonical ensemble. 
As we will see, this turns out to be useful to obtain an exact correspondence between the local operators and plane partitions and to study the asymptotic behaviors. 

We define a grand canonical index by
\begin{align}
\label{gindex}
\Xi_k(z;q)
&=1+\sum_{N=1}^{\infty} \mathcal{I}_{N,k}(q) q^{kN} z^N. 
\end{align}
where $z=e^{\mu}$ plays a role of the fugacity and $\mu$ is the chemical potential. 
Then the grand canonical potential is given by
\begin{align}
J_k(z;q)&=\log \Xi_k(z;q). 
\end{align}
We find that the grand canonical index (\ref{gindex}) is simply given by
\footnote{
We have numerically checked that (\ref{gindex_formula}) reproduces the canonical indices. 
It would be nice to prove this analytically. 
}
\begin{align}
\label{gindex_formula}
\Xi_{k}(z;q)&=\prod_{n=1}^{\infty}\prod_{m=1}^{k}\frac{1}{(1-z q^{kn+2(m-1)})^n}. 
\end{align}
For example, for $k=1$ the grand canonical index is 
\footnote{
The MacMahon function as the generating function for plane partitions coincides with 
the partition function of free conformally coupled scalar on $S^1\times S^2$ \cite{Nekrasov:2017cih}. 
The function (\ref{gindex_k1}) can be also understood as the partition function of free conformally coupled scalar 
where $z$ is the fugacity for its angular momentum. 
The author thanks Nikita Nekrasov for sharing his idea. 
}
\begin{align}
\label{gindex_k1}
\Xi_{1}(z;q)&=\prod_{n=1}^{\infty}\frac{1}{(1-z q^{n})^n}
\end{align}
and the grand potential is
\begin{align}
\label{gpot_k1}
J_1(z;q)&=\sum_{l=1}^{\infty}\frac{z^l}{l}\frac{q^l}{(1-q^l)^2}. 
\end{align}

From (\ref{gindex_k1}) one can check that the grand canonical index obeys a relation
\begin{align}
\label{gindex_prop1}
\Xi_{1}(zq^{-1};q)&=
\frac{\Xi_1(z;q)}{(z;q)_{\infty}}.
\end{align}
By using the identity \cite{MR2128719}
\begin{align}
\frac{1}{(z;q)_{\infty}}&=\sum_{n=0}^{\infty}\frac{z^n}{(q;q)_n}
\end{align}
for $|z|<1$ we get from (\ref{gindex_prop1}) a recursion relation
\begin{align}
\label{recursion}
\mathcal{I}_{N,1}(q)&=
\frac{1}{1-q^N}\sum_{n=0}^{N-1}
\frac{q^n}{(q;q)_{N-n}}\mathcal{I}_{n,1}(q). 
\end{align}

Here we observe that the expression (\ref{gindex_k1}) has a combinatorial interpretation 
as a generating function for the number of plane partitions. 
The R.H.S. of (\ref{gindex_k1}) is known as a generating function for the number $\beta(n,m)$ of plane partitions of $n$ with trace $m$ \cite{MR325407}. 
So we can write 
\begin{align}
\label{gindex_k1exp}
\Xi_{1}(z;q)&=\sum_{n=0}^{\infty}\sum_{m=0}^{\infty}\beta(n,m) q^n z^m. 
\end{align}

For example, the $z^2$ term in $\Xi_1(z;q)$ is 
\begin{align}
\label{gindex_k1_z2}
\frac{q^2+q^4}{(1-q)^2(1-q^2)^2}z^2&=
(q^2+2 q^3+6 q^4+10 q^5+19 q^6+28 q^7+44 q^8+\cdots)z^2.
\end{align}
While there are $Q(4)=13$ plane partitions of $4$
\begin{align}
&\young(4),& 
&\young(31),& 
&\young(22),& 
&\young(211),&
&\young(1111),\nonumber\\
&\young(3,1),& 
&\young(2,2),& 
&\young(21,1),&
&\young(11,11),&
&\young(111,1),\nonumber\\
&\young(11,1,1),&
&\young(2,1,1),&
&\young(1,1,1,1),
\end{align}
there are $\beta(4,2)=6$ plane partitions of $4$ with trace $2$
\begin{align}
\label{pp_4}
&\young(22),& 
&\young(211),& 
&\young(2,2),
\nonumber\\
&\young(21,1),&
&\young(11,11),&
&\young(2,1,1), 
\end{align} 
which correspond to the term $6q^4$ in (\ref{gindex_k1_z2}). 
Similarly, one can obtain from (\ref{gindex_k1_z2}) $\beta(5,2)=10$. 
This corresponds to the following $10$ plane partitions of $5$ with trace 2: 
\begin{align}
\label{pp_5}
&\young(2111),& 
&\young(221),& 
&\young(211,1),&
&\young(22,1),
\nonumber\\
&\young(111,11),&
&\young(2,1,1,1),&
&\young(2,2,1),&
&\young(21,1,1),&
\nonumber\\
&\young(21,2),& 
&\young(11,11,1).
\end{align}

Comparing (\ref{gindex}) and (\ref{gindex_k1exp}), 
we deduce a correspondence between 
the protected local operators parametrizing the geometry $\mathbb{C}^2$ probed by $N$ M2-branes and plane partitions with trace $N$. 
The scaling dimension $\Delta$ carried by the local operator translates into the norm $n$ $=$ $\sum_{i,j} n_{i,j}$ of the plane partitions. 
As will be explained, the vacuum $\nu$ in which the operator gets a vev 
and the flavor charge $f$ of the operator are also encoded in the plane partition. 
The correspondence is summarized as
\begin{align}
\label{corr}
\begin{array}{c|c|c}
\textrm{M2-brane SCFT operators}&\textrm{plane partitions}&\textrm{relation} \\ \hline 
\textrm{rank of gauge group $N$}& \textrm{trace $m$}&N=m \\
\textrm{scaling dimension $\Delta$}& \textrm{norm $n$}&\Delta=(n-N)/2 \\
\textrm{vacuum $\nu$}&\textrm{partition $\lambda$ of diagonal}&\nu=\lambda \\
\textrm{flavor charge $f$}&\textrm{$i$-trace $\tau_i$}&f=\sum_{i>0} \tau_i-\sum_{i<0}\tau_i \\
\end{array}
\end{align}

To illustrate the correspondence (\ref{corr}), we work out the previous example 
that appears from the $z^2$ term (\ref{gindex_k1_z2}). 
Alternatively, we get from (\ref{gindex_k1_z2}) the canonical index for $N=2$
\begin{align}
\label{index_k1_z2}
\mathcal{I}_{2,1}&=
1+2q+6q^2+10q^3+19q^4+28q^5+44q^6+\cdots
\end{align}
The term $6q^2$ can be found in the $U(2)_1\times U(2)_{-1}$ ABJM model as the gauge invariant polynomials in the twisted hypermultiplet $(T,\widetilde{T})$ dressing the monopole
\begin{align}
\label{abjm_z2a}
&v^{1,1;1,1}T^2,& 
&v^{2,0;2,0}T^2,&
&v^{-1,-1;-1,-1}\widetilde{T}^2
,\nonumber\\
&v^{1,-1;1,-1}T\widetilde{T},& 
&\Tr(T\widetilde{T}),& 
&v^{-2,0;-2,0}\widetilde{T}^2.
\end{align}
On the other hand, in the $U(2)$ ADHM theory with one flavor it counts the following six Coulomb branch operators with dimension $1$:
\begin{align}
\label{adhm_z2a}
&v^{1,1},& 
&v^{2,0},&
&v^{-1,-1}
,\nonumber\\
&v^{1,-1},& 
&\Tr \varphi,&  
&v^{-2,0}.
\end{align}
The operators (\ref{abjm_z2a}) in the ABJM theory and (\ref{adhm_z2a}) in the ADHM theory 
correspond to the plane partitions (\ref{pp_4}). 
Analogously, the term $10q^3$ in (\ref{index_k1_z2}) which corresponds to the $10$ plane partitions (\ref{pp_5}) of $5$ with trace $2$ 
counts the gauge invariant operators
\begin{align}
\label{abjm_z2b}
&v^{3,0;3,0}T^3,&
&v^{2,1;2,1}T^3,&
&v^{2,-1;2,-1}T^2\widetilde{T},& 
&v^{1,0;1,0}T^{(1)} (T^{(2)}\widetilde{T}),
\nonumber\\
&v^{1,0;1,0}T^{(2)} (T^{(1)}\widetilde{T}) ,&
&v^{-3,0;-3,0}\widetilde{T}^3,&
&v^{-2,-1;-2,-1}\widetilde{T}^3,& 
&v^{1,-2;1,-2} {\widetilde{T}}^2 T,
\nonumber\\
&v^{-1,0;-1,0}\widetilde{T}^{(1)} (T\widetilde{T}^{(2)}),& 
&v^{-1,0;-1,0}\widetilde{T}^{(2)} (T\widetilde{T}^{(1)})
\end{align}
in the ABJM theory and 
\begin{align}
\label{abjm_z2b}
&v^{3,0},&
&v^{2,1},&
&v^{2,-1},& 
&v^{1,0}\varphi^{(1)},
\nonumber\\
&v^{1,0}\varphi^{(2)},&
&v^{-3,0},&
&v^{-2,-1},& 
&v^{1,-2},
\nonumber\\
&v^{-1,0}\varphi^{(1)},& 
&v^{-1,0}\varphi^{(2)}
\end{align}
in the ADHM theory. 
Here we have introduced the superscripts of the scalar fields to label the distinct components 
which appear as the irreducible representation of the gauge group broken by the magnetic flux. 

We observe that 
a certain flavor charge $f$ carried by the corresponding operator for a given plane partition is encoded as the sum of the entries above the diagonal 
minus the sum of the entries below the diagonal. 
Let $\tau_i(\pi)$ be the $i$-trace \footnote{The $i$-trace is defined in \cite{MR607040}. The trace $\tau$ is viewed as the $0$-trace. } 
which is defined as the sum of the entries in the $i$-th diagonal of the plane partition $\pi$ 
where $i$-th diagonal is the sequence of all entries $n_{kl}$ with $i=l-k$. 
Then the flavor charge $f$ is given by 
\begin{align}
\label{f_charge}
f&=\sum_{i> 0}\tau_i(\pi) -\sum_{i< 0}\tau_i(\pi). 
\end{align}
The transpose of the plane partition $\pi=\{n_{ij}\}$ which is defined by $\pi^{*}=\{n_{ji}\}$ 
can be interpreted as a conjugation on the operator. 
For example, in the ABJM theory it flips signs of the GNO charges of monopoles and exchanges $T$ and $\widetilde{T}$. 
In the ADHM theory it just flips signs of the GNO charges. 

The plane partition whose entries below the diagonal are zero corresponds to the operator with the largest positive charge 
for a given dimension while the plane partition that has only the entries below the diagonal correspond to the operator with the largest negative charge. 
For example, the plane partitions $\small \young(2111)$ and $\small \young(221)$ whose entries below the diagonal vanish correspond to 
the dressed monopole operators $v^{3,0}T^3$ and $v^{2,1} T^3$ with the flavor charge $+3$ in the $U(2)_{1}\times U(2)_{-1}$ ABJM theory 
and the monopole operators $v^{3,0}$ and $v^{2,1}$ of the GNO charge $+3$ in the $U(2)$ ADHM theory with one flavor. 

A symmetric plane partition satisfying $n_{ij}=n_{ji}$ for all $i,j$ is invariant under the transpose. 
It is realized as a certain self-conjugate operator. 
In (\ref{pp_4}) there are two symmetric plane partitions $\small \young(21,1)$ and $\small \young(11,11)$. 
They are identified with the self-conjugate operators $v^{1,-1;1,-1}T\widetilde{T}$ and $\Tr (T\widetilde{T})$ in the ABJM theory 
and $v^{1,-1}$ and $\Tr \varphi$ in the ADHM theory. 

The plane partition that is not invariant under the transpose realizes the operator that is not self-conjugate.  
Such a non-symmetric plane partition can be constructed by adding boxes to a symmetric plane partition. 
For example, in the ABJM model, one can add $\small \young(1)$ and $\small \young(2)$ to a symmetric plane partition in the top row.  
Then the GNO charge of the corresponding monopole operator is increased by $(1,0)$ or $(1,1)$ so that it is dressed by $T$ or $T^2$ respectively. 
Similarly when $\small \young(1)$ or $\small \young(2)$ is added in the left column, 
the monopole operator acquires the GNO charge $(-1,0)$ or $(-1,-1)$ together with $\tilde{T}$ or $\widetilde{T}^2$ respectively. 

Recall that the canonical index can be also obtained from the formula (\ref{SymNC2}) in terms of the generating function (\ref{block_IR}) for column-strict plane partitions. 
The relation to the previous formula (\ref{SymNC2}) follows 
from the so-called Frobenius construction \cite{MR618067}, that is a bijection between plane partitions and pairs of column-strict plane partitions of the same shape. 

It is shown \cite{MR618067} that 
for a plane partition $\pi=\{n_{ij}\}$ of $n$ 
there exists a pair of two column-strict plane partitions $P=\{p_{ij}\}$ and $Q=\{q_{ij}\}$ of the shape $\lambda$ 
in such a way that 
\footnote{Another bijection between plane partitions and pairs of column-strict plane partitions of the same type is shown in \cite{MR299574}. 
This is realized as a \textit{conjugate} (called \textit{aspects} in \cite{MR2417935}) of this plane partition. }
\begin{align}
\label{Frobenius}
n_{ij}&=
\begin{cases}
| \{ k: p_{jk}\ge i-j+1 \} | &\textrm{for $j\le i$}\cr 
| \{ k: q_{ik}\ge j-i+1 \} | &\textrm{for $i\le j$}\cr 
\end{cases}
\end{align}
where $|S|$ is the cardinality of the set $S$. 
It follows that the shape $\lambda$ is identified with the diagonal of $\pi$ 
and that the largest part of $P$ (resp. $Q$) is the number of rows (resp. columns) in $\pi$. 
Also it is shown that $\pi$ is symmetric if and only if $P$ and $Q$ are equivalent. 

We observe that 
the shape $\lambda$ of the column-strict plane partitions $(P,Q)$ is identified with the massive vacuum $\nu$ appearing in the formula (\ref{SymNC2}). 
From the perspective of the plane partition $\pi$ it is the partition $\lambda$ of the $0$-th diagonal. 

For example, 
for the plane partitions (\ref{pp_4}) of $4$ with trace $2$ we have 
the pairs $(P,Q)$ of column-strict plane partitions 
\begin{align}
\label{pp_4_PQ}
&\begin{array}{c||c|c}
\pi&P&Q\\ \hline \hline 
\small \young(22)&\small \young(11)&\small \young(22)\\ 
&& \\
\small \young(211)&\small \young(11)&\small \young(31)\\
&& \\
\small \young(2,2)&\small \young(22)&\small \young(11)\\
\end{array}& 
&\begin{array}{c||c|c}
\pi&P&Q\\ \hline \hline 
\small \young(21,1)&\small \young(21)&\small \young(21)\\ 
&& \\
\small \young(11,11)&\small \young(2,1)&\small \young(2,1)\\
&& \\
\small \young(2,1,1)&\small \young(31)&\small \young(11)\\
\end{array}
\end{align}
and for the plane partitions (\ref{pp_5}) of $5$ with trace $2$ we have 
\begin{align}
\label{pp_5_PQ}
&\begin{array}{c||c|c}
\pi&P&Q\\ \hline \hline 
\small \young(2111)&\small \young(11)&\small \young(41)\\ 
&& \\
\small \young(221)&\small \young(11)&\small \young(32)\\
&& \\
\small \young(211,1)&\small \young(21)&\small \young(31)\\
&& \\
\small \young(22,1)&\small \young(21)&\small \young(22)\\
&& \\
\small \young(111,11)&\small \young(2,1)&\small \young(3,1)\\
\end{array}& 
& 
\begin{array}{c||c|c}
\pi&P&Q\\ \hline \hline 
\small \young(2,1,1,1)&\small \young(41)&\small \young(11)\\ 
&& \\
\small \young(2,2,1)&\small \young(32)&\small \young(11)\\
&& \\
\small \young(21,1,1)&\small \young(31)&\small \young(21)\\
&& \\
\end{array},& 
&\begin{array}{c||c|c}
\pi&P&Q\\ \hline \hline 
\small \young(21,2)&\small \young(22)&\small \young(21)\\
&& \\
\young(11,11,1)&\young(3,1)&\young(2,1)\\
\end{array}
\end{align}
While in (\ref{pp_4_PQ}) there are $5$ plane partitions characterized by the shapes $\tiny \yng(2)$ of $(P,Q)$ 
and a single plane partition characterized by $\tiny \yng(1,1)$ of $(P,Q)$, 
in (\ref{pp_5_PQ}) we have $8$ plane partitions associated to $\tiny \yng(2)$ 
and $2$ plane partitions associated to $\tiny \yng(1,1)$. 

On the other hand, from the formula (\ref{SymNC2}) we can write the canonical index (\ref{index_k1_z2}) as
\begin{align}
\mathcal{I}_{2,1}(q)&=
\chi_{\tiny \yng(2)}(q)^2+\chi_{\tiny \yng(1,1)}(q)^2
\end{align}
where 
\begin{align}
\label{block_U2a}
\chi_{\tiny \yng(2)}(q)^2&
=\frac{1}{(1-q)^2 (1-q^2)^2}
\nonumber\\
&=1+2q+5q^2+8q^3+14q^4+\cdots,\\
\label{block_U2b}
\chi_{\tiny \yng(1,1)}(q)^2
&=\frac{q^2}{(1-q)^2 (1-q^2)^2}
\nonumber\\
&=q^2+2q^3+5q^4+8q^5+14q^6+\cdots.
\end{align}
The terms $5q^2$ and $8q^3$ in (\ref{block_U2a}) count the plane partitions labeled by the shape $\tiny \yng(2)$ of $(P,Q)$ 
in (\ref{pp_4_PQ}) and (\ref{pp_5_PQ}) respectively. 
Also the terms $q^2$ and $2q^3$ in (\ref{block_U2b}) count those for $\tiny \yng(1,1)$ 
in (\ref{pp_4_PQ}) and (\ref{pp_5_PQ}) respectively. 
In this way one can explicitly check that 
the shape $\lambda$ of the pair $(P,Q)$ of the column-strict plane partitions obtained from the Frobenius construction (\ref{Frobenius}) 
is identified with the massive vacuum $\nu$ in the formula (\ref{SymNC2}). 

Hence we also have the correspondence between 
the local operators and the pairs $(P,Q)$ of column-strict plane partitions. 
Let $|P|$ and $|Q|$ be the norms of the column-strict plane partitions $P$ and $Q$ respectively. 
Let $\lambda$ be the shape of $P$ and $Q$. 
From the point of view of the column-strict plane partitions, 
the number $N$ of M2-branes and the isolated massive vacuum $\nu$ are encoded by 
the weight $|\lambda|$ $=$ $\sum_i \lambda_i$ and the shape $\lambda$. 
The sum $|P|+|Q|$ is equal to the twice the norm $n$ of $\pi$, 
whereas the difference $|Q|-|P|$ encodes the flavor charge $f$ of the corresponding operator
\begin{align}
|Q|+|P|&=2n=2N+2\Delta,\\
|Q|-|P|&=f. 
\end{align}
An exchange of $P$ with $Q$ corresponds to the conjugation on the operator. 
The correspondence between the local operators and the pairs of colum-strict plane partitions is given by
\begin{align}
\label{corr2}
\begin{array}{c|c|c}
\textrm{M2-brane SCFT operators}&\textrm{  column-strict p.p.s  }&\textrm{relation} \\ \hline 
\textrm{rank of gauge group $N$}& \textrm{weight $|\lambda|$} &N=|\lambda| \\
\textrm{scaling dimension $\Delta$}& \textrm{norms $|P|, |Q|$}&\Delta=(|Q|+|P|-2N)/2 \\
\textrm{vacuum $\nu$}&\textrm{shape $\lambda$}&\nu=\lambda \\
\textrm{flavor charge $f$}&\textrm{norms $|P|, |Q|$}&f=|Q|-|P| \\
\end{array}
\end{align}

Therefore, 
according to the correspondence (\ref{corr}) or (\ref{corr2}), the operator counting problem is translated into the enumeration of the plane partitions. 
There are ten symmetry operations on plane partitions and the problem of enumerating plane partitions with the symmetries \cite{MR859302}. 
It would be interesting to find further physical implications of these symmetries and give the holographic interpretation of the correspondence between the operator and the plane partition. 

\subsection{High-temperature limit}
The $N^{3/2}$ behavior \cite{Klebanov:1996un} of the number of degrees of freedom of $N$ M2-branes is obtained from 
the large $N$ analysis of sphere partition functions 
\cite{Drukker:2010nc,Drukker:2011zy,Fuji:2011km,Marino:2011eh,
Herzog:2010hf,Santamaria:2010dm,Martelli:2011qj,Cheon:2011vi,Jafferis:2011zi,Gabella:2011sg}, 
twisted indices \cite{Benini:2015eyy,Benini:2016hjo,Azzurli:2017kxo,Liu:2017vbl} and full-superconformal indices \cite{Choi:2019zpz}. 
Here we set $q=e^{-\beta}$ and study the high-temperature limit $\beta\rightarrow 0$ of the grand potential. 
We write the grand potential (\ref{gpot_k1}) as
\begin{align}
\label{gpot_k1_1}
J_1(z;q)&=\beta\sum_{l=1}^{\infty}z^l \frac{q^l}{l\beta (1-q^l)^2}
=\beta \sum_{l=1}^{\infty}z^l f(l\beta)
\end{align}
where
\begin{align}
\label{summand_f}
f(x)&:=\frac{e^{-x}}{x(1-e^{-x})^2}. 
\end{align}
Note that the function (\ref{summand_f}) can be expanded as
\begin{align}
f(x)&=-\sum_{n=-3}^{\infty} 
\frac{(n+2)B_{n+3}}{(n+3)!}z^n
\end{align}
where $B_{n}$ are the Bernoulli numbers. 
We consider the asymptotic property when the fugacity $z$ is near roots of unity. 
Let $z$ be a primitive $b$-th root of unity
\begin{align}
\label{root_unity}
z&=z_{ab}:=e^{\frac{2\pi ia}{b}}.
\end{align}
Then the sum in (\ref{gpot_k1_1}) is performed by replacing $l$ with 
$bl+j$, with $l=0,1,2,\cdots$ and $j=1,\cdots, b$. 
So we have
\begin{align}
\label{gpot_k1_2}
J_1(z;q)&=\beta \sum_{j=1}^{b} z_{ab}^j \sum_{l=0}^{\infty} f\left( \left(l+\frac{j}{b}\right) b\beta \right). 
\end{align}
The asymptotic behavior of (\ref{gpot_k1_2}) can be obtained from the following generalized Euler-Maclaurin summation formula \cite{bringmann2021distributions}: 
\begin{thr}
If $f(x)$ has the asymptotic expansion 
\begin{align}
f(x)&=\sum_{n=n_{0}}^{\infty}c_n x^n
\end{align}
for $n_0\in \mathbb{Z}$ in the domain $D_{\theta}=\{x=re^{i\alpha}:r\ge 0,|\alpha|\le\theta\}$ as $x\rightarrow 0$, 
then we have
\begin{align}
\sum_{n=0}^{\infty}
f\left( (n+a)x \right)
&\sim 
\sum_{n=n_0}^{-2} c_n \zeta(-n,a)x^n
+\frac{I^{*}_{f,A}}{x}
-\frac{c_{-1}}{x}\left( \log(Ax)+\psi(a)+\gamma \right)
\nonumber\\
&-\sum_{n=0}^{\infty} c_n \frac{B_{n+1}(a)}{n+1}x^n
\end{align}
as $x\rightarrow 0$ for $0<a\le 1$ and  some $A\in \mathbb{R}_{+}$. 
Here $\zeta(s,z)$ $:=$ $\sum_{n=0}^{\infty}1/(n+z)^s$ is the Hurwitz zeta function, 
$\psi(x)$ $:=$ $\Gamma'(x)/\Gamma(x)$ is the digamma function, 
$\gamma$ is the Euler–Mascheroni constant, 
$B_n(x)$ are the Bernoulli polynomials 
and 
\begin{align}
I_{f,A}^{*}&=\int_{0}^{\infty}du 
\left[
f(u)-\sum_{n=n_0}^{-2} c_n u^n
-\frac{c_{-1}e^{-Au}}{u}
\right]. 
\end{align}
\end{thr}
Applying Theorem 2 to (\ref{gpot_k1_2}), we find 
\footnote{
The same result is also obtained in \cite{cesana2021asymptotic}. 
}
\begin{align}
\label{gpot_k1_3}
J_1(z;q)
&\sim 
\beta\sum_{j=1}^{b}z_{ab}^j 
\Biggl[
\frac{\zeta(3,\frac{j}{b})}{(b\beta)^3}
+\frac{I_{F,A}^{*}}{b\beta}
+\frac{1}{12b\beta}
\left(
\log(Ab\beta)+\psi(\frac{j}{b})+\gamma
\right)
\nonumber\\
&+\sum_{n=0}^{\infty}
\frac{(n+2)B_{n+3}}{(n+1)(n+3)!} B_{n+1}(\frac{j}{b})(b\beta)^n
\Biggr]
\nonumber\\
&=
\sum_{j=1}^{b}z_{ab}^j \frac{\zeta(3,\frac{j}{b})}{b^3\beta^2}
+\frac{1}{12b}\sum_{j=1}^{b}z_{ab}^j\psi(\frac{j}{b})+\mathcal{O}(|\beta|)
\nonumber\\
&=
\frac{\mathrm{Li}_{3}(z_{ab})}{\beta^2}
+\frac{1}{12}\log (1-z_{ab})+\mathcal{O}(|\beta|)
\qquad \textrm{as $\beta\rightarrow 0$}
\end{align}
where the first equality follows from the fact that the sum of roots of unity vanishes 
and the second equality is obtained from the relation \cite{110}
\begin{align}
\sum_{j=1}^{b} z_{ab}^j \zeta(3,\frac{j}{b})&=\sum_{j=1}^{b}z_{ab}^j\sum_{n=0}^{\infty}\frac{b^3}{(bn+j)^3}
=b^3\mathrm{Li_{3}}(z_{ab})
\end{align}
where $\mathrm{Li}_{p}(z)=\sum_{k=1}^{\infty} z^k/k^p$ for $|z|<1$ is the polylogarithm function and the relation \cite{bringmann2021distributions}
\begin{align}
\sum_{j=1}^{b}z_{ab}^j\psi(\frac{j}{b})&=b\log(1-z_{ab}). 
\end{align}
We remark that the leading trilogarithm $\mathrm{Li}_{3}(z)$ in the grand potential (\ref{gpot_k1_3}) also appears in the grand potential for the sphere partition function of the ABJM theory \cite{Marino:2011eh} and the ADHM theory \cite{Mezei:2013gqa,Grassi:2014vwa,Hatsuda:2014vsa}. 
It is crucial for the $N^{3/2}$ growth of the free energy. 
It would be interesting to investigate the phase diagram by analyzing more details of the asymptotic behavior upon varying the chemical potential. 

\subsection{Large $N$ scaling dimension}
Since the grand canonical index (\ref{gindex_k1}) for $k=1$ is a generating function for plane partitions of $n$ with trace $N$, 
the combinatorics can give an alternative approach to the asymptotic behavior of the index. 

The asymptotics of plane partitions can be studied in the theory of random plane partitions 
by assigning the probability $1/Q(n)$ for each partition of $n$ and introducing the uniform probability measure $\mathbb{P}$ on the set of plane partitions of $n$. 
Let $\tau(n)$ be the trace of a plane partition of $n$. 
In terms of the probability measure, we can write the generating function (\ref{gindex_k1exp}) or equivalently the grand canonical index (\ref{gindex_k1}) as
\begin{align}
\label{gindex_k1prob}
\Xi_1(z;q)&=1+\sum_{n=1}^{\infty} Q(n) q^n \sum_{N=1}^n \mathbb{P}(\tau(n)=N)z^N
\nonumber\\
&=1+\sum_{n=1}^{\infty} Q(n) \varphi_n(z)q^n
\end{align}
where 
\begin{align}
\label{prob_gene}
\varphi_n(z)&=\sum_{N=1}^{n}\mathbb{P}(\tau(n)=N) z^N
\end{align}
is a probability generating function. 
Since we can get an expectation value $\langle N\rangle$ of trace $\tau(n)$ of plane partitions of $n$ 
with respect to the probability measure $\mathbb{P}$ 
by differentiating $\varphi_n(z)$ with respect to $z$
\begin{align}
\langle N\rangle(n)&=
\frac{d\varphi_n(z)}{dz}\Bigl|_{z=1}=\sum_{N=1}^n N \mathbb{P}(\tau(n)=N),
\end{align}
we get from (\ref{gindex_k1prob})
\begin{align}
\label{d_Xi1}
\frac{\partial}{\partial z}\Xi_1(z;q)\Bigl|_{z=1}&=
\sum_{n=1}Q(n)\langle N\rangle(n) q^n. 
\end{align}
On the other hand, from (\ref{gindex_k1}) we have 
\begin{align}
\label{d_Xi2}
\frac{\partial}{\partial z}\Xi_1(z;q)\Bigl|_{z=1}
&=M(q)\prod_{n=1}^{\infty} \frac{nq^n}{1-q^n}
\end{align}
where $M(q)$ is the MacMahon function (\ref{MacMahon}). 
We can obtain the mean value $\langle N\rangle(n)$ 
by expanding (\ref{d_Xi2}) with respect to $q$ and comparing it with (\ref{d_Xi1}). 
The asymptotics can be obtained from the following theorem \cite{MR3785797}: 

\begin{thr}
Let 
\begin{align}
M(q)F(q)&=\sum_{n=0}^{\infty} C_n q^n
\end{align}
where $M(q)$ is the MacMahon function and $F(q)$ is a function which satisfies 
\begin{enumerate}[label=(\roman*)]

\item Condition
\begin{align}
\label{cond_1}
\lim_{n\rightarrow \infty} \frac{F(r_n e^{i\theta})}{F(r_n)}=1 \qquad \textrm{for $|\theta|\le \delta(r_n)$}
\end{align}
where $r_n$ is a solution to the equation $rM'(r)/M(r)=n$ 
and $\delta(r)$ is some function defined over $r\in (R_0,\rho) \subset (0,\rho)$ for some $\rho$ in which $M(r)>0$. 

\item Condition
\begin{align}
\label{cond_2}
F(q)&=\mathcal{O}(e^{C/(1-|q|)^D})
\end{align}
for some $C>0$ and $D\in (0,2/3)$ as $|q|\rightarrow 1$. 

\end{enumerate}
Then we have 
\begin{align}
\frac{C_n}{Q(n)}&=F(e^{-d_n})\left(1+\mathcal{O}(1) \right)+\mathcal{O}(e^{-cn^{2/9}/\log^2 n})
\end{align}
as $n\rightarrow \infty$. 
Here $\{d_n\}_{n\ge 1}$ is a sequence with the following expansion as $n\rightarrow \infty$
\begin{align}
d_n&=\left( \frac{2\zeta(3)}{n} \right)^{1/3}
-\frac{1}{36n}+\cdots
\end{align}
and $c>0$ is some constant. 
\end{thr}
The proof of Theorem 3 can be found in \cite{MR3785797} where Hayman's theorem \cite{MR80749} and the Cauchy integral are employed. 
For our purpose we take $F(q)$ as $\prod_{n}nq^n/(1-q^n)$ in (\ref{d_Xi2}). 
\footnote{It can be shown that $F(q)$ $=$ $\prod_n n q^n/(1-q^n)$ satisfies the conditions (\ref{cond_1}) and (\ref{cond_2}). See \cite{MR3785797}.}
Applying Theorem 3 we get
\begin{align}
\label{N_large_n}
\langle N\rangle(n)
&\sim F(e^{-d_n})
=d_n^{-2}\sum_{k=1}^{\infty}\frac{kd_n e^{-kd_n}}{1-e^{-kd_n}}d_n
\nonumber\\
&\sim d_n^{-2}\int_{0}^{\infty} \frac{x}{e^x-1}dx=d_n^{-2}\zeta(2)
=\frac{\pi^2}{6} \left( \frac{n}{2\zeta(3)} \right)^{2/3} 
\qquad \textrm{as $n\rightarrow \infty$}. 
\end{align}
Since (\ref{corr}) relates the norm $n$ of plane partitions to the scaling dimension $\Delta$ carried by the local operators, 
(\ref{N_large_n}) gives rise to the asymptotic behavior of the large scaling dimension
\begin{align}
\label{largeD}
\Delta&\sim \frac{\zeta(3) 6^{3/2}}{\pi^3}N^{3/2}. 
\end{align}
For example, from the perspective of the $U(N)$ ADHM theory with one flavor, 
(\ref{largeD}) would imply that the scaling dimension, i.e. the canonical R-charge of the Coulomb branch operators, follows the $N^{3/2}$ growth in the large $N$ limit. 
The growth (\ref{largeD}) of the large scaling dimension is expected to characterize the local operators in the 3d SCFT of $N$ M2-branes moving in $\mathbb{C}^2$. 

\subsection*{Acknowledgements}
The author would like to thank 
Yasuyuki Hatsuda, Hirotaka Hayashi, Kimyeong Lee, Sungjay Lee, Nikita Nekrasov, Tomoki Nosaka and Douglas J. Smith for useful discussions and comments. 
This work was supported by KIAS Individual Grants (PG084301) at Korea Institute for Advanced Study. 

\bibliographystyle{utphys}
\bibliography{ref}

\end{document}